\documentstyle[11pt]{article}
\def\v1{\vspace{1cm}}
\def\be{\begin{equation}}
\def\ee{\end{equation}}
\def\bc{\begin{center}}
\def\ec{\end{center}}
\def\ik{\partial}
\def\vh{\varphi}

\newcommand{\bea}{\begin{eqnarray}}
\newcommand{\eea}{\end{eqnarray}}

\begin{document}

\title
{\bf Reparametrization-Invariant Path Integral in GR and "Big Bang"
of Quantum Universe}
\author{
M. Pawlowski,\\
{\normalsize\it Soltan Institute for Nuclear Studies} \\
{\normalsize\it Warsaw,Poland.}\\[0.3cm]
V.N. Pervushin \\
{\normalsize\it Joint Institute for Nuclear Research},\\
 {\normalsize\it 141980, Dubna, Russia.}\\
}

\date{\empty}

\maketitle

\medskip
PACS number(s):04.60.-m, 04.20.Cv, 98.80.Hw (Quantum Gravity)
\medskip


\begin{abstract}
{\small
{
The reparametrization-invariant
generating functional for the unitary and causal perturbation theory
in general relativity in a finite space-time is obtained.
The region of validity of
the Faddeev-Popov-DeWitt functional is studied.
It is shown that the invariant content of general relativity as a
constrained system can be covered by two "equivalent" unconstrained systems:
the "dynamic" (with "dynamic" evolution parameter as the metric scale factor)
and "geometric" (given by the
Levi-Civita type canonical transformation to the action-angle variables
where the energy constraint converts into a new momentum).

"Big Bang", the Hubble evolution, and creation of matter fields by
the "geometric" vacuum are described by the inverted Levi-Civita
transformation of the geomeric system into the dynamic one.
The particular case of the Levi-Civita transformations are the Bogoliubov ones
of the particle variables (diagonalizing the dynamic Hamiltonian) to
the quasiparticles (diagonalizing the equations of motion).
 The choice of
initial conditions for the "Big Bang" in the form of the Bogoliubov
(squeezed) vacuum reproduces all stages of the evolution
of the Friedmann-Robertson-Walker universe in their conformal
(Hoyle-Narlikar) versions.
}}

\end{abstract}



\section{Introduction}

"Big Bang" as the beginning of the Hubble evolution of a universe
is described as a pure classical phenomenon on the basic of particular
solutions of the Einstein equations in general relativity in the
homogeneous approximation.
A strange situation consists in that the highest level
of the theory, i.e., the Faddeev-Popov-DeWitt generating functional
for unitary S-matrix~\cite{fp2,dw}, neglects the questions
about the evolution of a universe which are in the competence of
the simplest classical approximation.
There is an opinion that the solution of the "Big Bang" problem
in quantum theory goes beyond the scope  of
the unitary perturbation theory and even of general relativity. To answer
these questions, we need a more general theory of the type of
superstring~\cite{gsw}.

According to another point of view, the reason of theoretical difficulties
in understanding the "Big Bang" phenomenon is not the Einstein theory, but
the non-invariant method of its quantization. In particular,
for the Faddeev-Popov-DeWitt unitary S-matrix the non-invariant
coordinate time is considered as the time of evolution, whereas
an observer in a universe can observe and measure only invariants
of group of diffeomorphisms of the Hamiltonian dynamics, which includes
reparametrizations of the coordinate time~\cite{kpp,grg,plb,ps1}.

In the present paper we try to construct the unitary S-matrix
for general relativity in a finite space-time in terms of
the reparametrization-invariant evolution parameter, and
to answer the questions: What do Quantum Universe and Quantum Gravity mean?
What is the status of the "Big Bang" evolution in quantum theory?
What does creation of Quantum Universe mean?
on the level of perturbation theory,
using the scheme of the time-reparametrization-invariant Hamiltonian
reduction~\cite{grg}.

In the context of the Dirac generalized Hamiltonian theory for constrained systems
\cite{d1,d2,kp,gt}, this scheme means the explicit resolving of the first
class constraints to determine the constraint-shell action directly
in terms of invariants. In other words, we use the invariant reduction
of the action (to obtain an equivalent unconstrained system)
instead of the generally accepted non-invariant reduction
of the phase space by fixing gauges~\cite{fp2,dw,fp1} (see Fig. 1).

The example of the application of such an invariant reduction of the action
is the Dirac formulation of QED~\cite{cj}
directly in terms of the gauge-invariant
(dressed) fields as the proof of the adequateness of the Coulomb gauge with
the invariant content of classical equations.
Recall that the invariant reduction of the action is the way to obtain
the unconstrained Feynman integral~\cite{fad} for the foundation
of the intuitive Faddeev-Popov functional integral in gauges
theories~\cite{fp1}
and to reveal  collective excitations of the gauge fields in
the form of zero-modes of the first class constraints~\cite{gip} (see Fig. 2).

A constructive idea of
the considered invariant Hamiltonian reduction of general relativity
is to introduce
{\it the dynamic evolution parameter} as the
zero-mode collective excitation of metric
\cite{kpp,grg,plb,ps1,Y,kuchar,torre,yaf}.
This {\it dynamic evolution parameter} can be identified with
the zero-Fourier harmonic of the space-metric
determinant ~\cite{grg,plb} (treated in cosmology as the cosmic scale factor),
whereas its conjugate momentum, i.e, the second (external) form,
plays the role of the localizable Hamiltonian of evolution.

The separation of this zero-mode evolution parameter
on the level of the action allows us to determine also the
{\it invariant geometric time}
formed by averaging  the time-like component
of a metric over the space coordinates~\cite{grg,plb,ps1}.

An observer reveals
the evolution of the universe (with "Big Bang") as
the dependence of the dynamic evolution parameter (i.e. cosmic scale factor)
on the geometric time.

The evolution of both a classical and a quantum universes
in terms of the geometric time has the form
of the canonical transformations~\cite{lc,sh,gkp} of the initial dynamic variables
to a new set of variables for which the total energy constraint becomes
a new momentum, and its conjugate variable (i.e., a new dynamic evolution parameter)
coincides with the geometric time onto equations of motion.

The content of the paper is the following.
Section 2 is devoted to the reparametrization-invariant Hamiltonian reduction
of general relativity. In section 3 we construct the generating
functional for the unitary
perturbation theory which includes "Big Bang" and Hubble evolution.
In Section 4, "Big Bang"  and Hubble evolution are reproduced
in lowest order of perturbation theory. In Section 5, we research the conditions
of validity of the conventional quantum field theory in
the infinite space-time limit.

\section{Reparametrization-invariant Hamiltonian dynamics of GR}

\subsection{Action and variables}

General relativity (GR) is
given by the singular Einstein-Hilbert action with the matter fields
\be
\label{gr}
W(g|\mu)= \int d^4x\sqrt{-g}[-\frac{\mu^2}{6} R(g)+{\cal L}_f] ~~~
~~\left(~\mu^2=M^2_{Planck}\frac{3}{8\pi}~\right)
\ee
and by a measurable interval
\be \label{dse1}
  (ds)^2_e=g_{\alpha\beta}dx^\alpha dx^\beta~.
\ee
They are invariant with respect to general coordinate transformations
\be \label{x}
x_{\mu} \rightarrow  x_{\mu}'=x_{\mu}'(x_{0},x_{1},x_{2},x_{3}).
\ee
The generalized Hamiltonian approach to GR was formulated by Dirac
and Arnovit, Deser and Misner \cite{ADM} as a theory of system with
constraints in $3+1$ foliated space-time
\be
\label{dse}
  (ds)^2=g_{\mu\nu}dx^\mu dx^\nu= N^2 dt^2-{}^{(3)}g_{ij}\breve{dx}{}^i
  \breve{dx}{}^j\;~~~~~\;\;(\breve{dx}{}^i=dx^i+N^idt)
\ee
with the lapse function $N(t,\vec x)$, three shift vectors $N^i(t,\vec x)$,
and six space components ${}^{(3)}g_{ij}(t,\vec x)$ depending on the
coordinate time $t$ and the space coordinates $\vec x$.
The Dirac-ADM parametrization of metric~(\ref{dse}) characterizes
a family of hypersurfaces $t=\rm{const.}$ with the unit normal vector
$\nu^{\alpha}=(1/N,-N^k/N)$ to a hypersurface and with the second
(external) form
\be \label{ext}
\frac{1}{N}({}^{(3)}\dot g_{ij})-\Delta_i N_j -\Delta_j N_i)
\ee
that shows how this hypersurface is embedded into the four-dimensional
space-time.

Coordinate transformations conserving the family of
hypersurfaces $t=\rm{const.}$
\be \label{gt}
t \rightarrow \tilde t=\tilde t(t);~~~~~
x_{i} \rightarrow  \tilde x_{i}=\tilde x_{i}(t,x_{1},x_{2},x_{3})~,
\ee
\be \label{kine}
\tilde N = N \frac{dt}{d\tilde t};~~~~\tilde N^k=N^i
\frac{\partial \tilde x^k }{\partial x_i}\frac{dt}{d\tilde t} -
\frac{\partial \tilde x^k }{\partial x_i}
\frac{\partial x^i}{\partial \tilde t}
\ee
are called a kinemetric subgroup of the group of general coordinate
transformations (\ref{x})~\cite{grg,vlad}.
The group of kinemetric transformations
is the group of diffeomorphisms of the generalized Hamiltonian dynamics.
It includes  reparametrizations of the nonobservable
time coordinate $\tilde t(t)$)~(\ref{gt})
that play the principal role in the procedure of the
reparametrization-invariant reduction discussed in the previous Sections.
The main assertion of the invariant reduction is the following:
the dynamic evolution parameter is not the coordinate but the variable
with a negative contribution to the energy constraint.
(Recall that this reduction is based on the explicit resolving of
the global energy constraint with respect to the conjugate momentum
of the dynamic evolution parameter to convert this momentum into
the Hamiltonian of evolution of the reduced system.)

A negative contribution to the energy constraint is given by
the space-metric-determinant logarithm. Therefore, following papers
\cite{kuchar,grg,torre,kpp,Y,yaf} we introduce an invariant evolution
parameter $\vh_0(t)$ as the zero Fourier harmonic component of this
logarithm (treated, in cosmology, as the cosmic scale factor).
This variable is distinguished in general relativity
by the Lichnerowicz conformal-type transformation of field
variables $f$ with the conformal weight $(n)$
\be \label{lich}
{}^{(n)}\bar f={}^{(n)}f \left(\frac{\vh_0(t)}{\mu}\right)^{-n}~,
\ee
where  $n=2,\;0,\;-3/2,\;-1$ for the tensor, vector,
spinor, and scalar fields, respectively, $\bar f$ is
so-called conformal-invariant variable used in GR
for the analysis of initial data~\cite{Y,L}.
In particular, for metric we get
\be \label{conf}
g_{\mu\nu}(t,\vec x)=
\left({\vh_0(t)\over \mu}\right)^2
\bar g_{\mu\nu}(t,\vec x)~\Rightarrow~
  (ds)^2=
\left({\vh_0(t)\over \mu}\right)^2
[\bar N^2 dt^2-{}^{(3)}\bar g_{ij}\breve{dx}{}^i \breve{dx}{}^j]~.
\ee
As the zero Fourier harmonic is extracted from the space metric determinant
logarithm, the space metric $\bar g_{ij}(t,\vec x)$ should be defined in
a class of nonzero harmonics
\be \label{gbar}
\int d^3x\log ||\bar g_{ij}(t,\vec x)||=0.
\ee
The transformational properties of the curvature $R(g)$ with respect to
the transformations~(\ref{conf}) lead to the action~(\ref{gr})
in the form~\cite{grg}
\be \label{Econf}
W(g|\mu)=W(\bar g|\vh_0) -\int\limits_{t_1 }^{t_2 }dt
\int\limits_{V_0}d^3x\vh_0
\frac{d}{dt}
(\frac{\dot \vh_0 \sqrt{\bar g}}{ \bar N}).
\ee
This form define the global lapse function $N_0$ as the average of the lapse
function $\bar N$ in the metric $\bar g$ over the kinemetric invariant space
volume
\be  \label{aven}
N_0(t)=\frac{V_0}{\int\limits_{V_0} d^3x \frac{\sqrt{\bar g(t, \vec x}}
{ \bar N(t, \vec x)}}~,~~~~~~~~~~~ \bar g=det({}^{(3)}\bar g)~,
~~~~~~~~~V_0=\int\limits_{V_0 }^{ }d^3x~,
\ee
where $V_0$ is a free parameter which in the perturbation theory
has the meaning of a finite volume of the free coordinate space.
The lapse function $\bar N(t, \vec x)$ can be factorized into the global
component $N_0(t)$ and the local one ${\cal N}(t, \vec x)$
\be \label{ncal1}
\bar N(t, \vec x) \bar g^{-1/2}:= N_0(t){\cal N}(t, \vec x):=N_q~,
\ee
where $\cal N$ fulfills normalization condition:
\be \label{ncal}
I[{\cal N}]:={1\over V_0}\int{d^3x\over {\cal N}}=1
\ee
that is imposed after the procedure of variation of action, to reproduce
equations of motion of the initial theory.
In the Dirac harmonical variables~\cite{d1} chosen as
\be \label{hv}
q^{ik}=\bar g \bar g^{ik},
\ee
the metric~(\ref{dse}) takes the form
\be \label{dse11}
(ds)^2=
\frac{\vh_0(t)^2}{\mu^2}
q^{1/2} \left(~N_q^2 dt^2 -q_{ij}\breve{dx}{}^i
  \breve{dx}{}^j \right),
~~~~~~~~ (q=det ( q^{ij} )) .
\ee
The Dirac-Bergmann version of action (\ref{Econf}) in terms of the introduced
above variables reads
~\cite{grg,plb}
\be \label{Egrc2}
W=
\int\limits_{t_1}^{t_2} dt \left\{
L+\frac{1}{2}\partial _t(P_0\vh_0)
\right\},
\ee
\be \label{Egrc3}
 L =
 \left[\int\limits_{V_0} d^3x
\left(\sum\limits_{F }^{ }P_F\dot F - N^i{\cal P}_i\right)\right] -
P_0 \dot\vh_0 - N_0\left[-\frac{P_0^2}{4V_0}+ I^{-1} H(\vh_0)\right]~,
\ee
where
\be \label{sum}
\sum\limits_{F }^{ }P_F\dot F = \sum\limits_{f }^{ }p_f\dot f-
\pi_{ij}\dot { q}^{ij};
\ee
\be \label{fpt1}
H(\vh_0)=\int d^3x{\cal N}{\cal H}(\vh_0)
\ee
is the total Hamiltonian of the local degrees of freedom,
\be \label{hh}
{\cal H}(\vh_0)=\frac{6}{\vh_0^2} q^{ij} q^{kl}[\pi_{ik}\pi_{jl}-
\pi_{ij}\pi_{kl}]
+\frac{\vh_0^2 q^{1/2}}{6}{}^{(3)}R(\bar g) +{\cal H}_f~,
\ee
and
\be \label{ph}
{\cal P}_i
=2[\nabla_k(q^{kl}\pi_{il})-\nabla_{i}(q^{kl}\pi_{kl})] +{\cal P}_{i f}~
\ee
are the densities of energy and momentum
and ${\cal H}_f,{\cal P}_f$ are contributions of the matter fields.
In the following, we call the set of the field variables $F$~(\ref{sum})
with the dynamic evolution parameter $\vh_0$ the field world space.
The local part of the momentum of the space metric determinant
\be \label{cs1}
\pi (t,x) :=  q^{ij}\pi_{ij}
\ee
is given in the class of functions with the non-zero Fourier harmonics,
so that
\be \label{cstr2}
\int d^3x  \pi (t,x)=0~.
\ee

The geometric foundation of introducting
the global variable ~(\ref{conf}) in GR was given in~\cite{yaf} as
the assertion about the nonzero value of the second form in the whole
space. This assertion (which contradicts the Dirac gauge $\pi=0$) follows
from the global energy constraint, as, in the lowest order of
the Dirac perturbation theory,
positive contributions of particle-like excitations to the
zero Fourier harmonic of the energy constraint can be compensated
only by the nonzero value of the second form.

The aim of this paper is to obtain the dynamic "equivalent" unconstrained
system in the field world space ($F,\vh_0$) by explicit
resolving the global energy constraint and to study the possibility
of the Hamiltonian description of the "equivalent"  unconstrained
system in terms of the reparametrization-invariant evolution
parameter $T$ defined by equation $N_0dt=dT$.

\subsection{Local constraints and equations of motion}

Following Dirac~\cite{d1} we formulate generalized Hamiltonian
dynamics for the considered system (\ref{Egrc2}). It means the inclusion
of momenta for ${\cal N}$ and $N_i$ and appropriate terms with Lagrange
multipliers
\be \label{multipl}
W^D=
\int\limits_{t_1}^{t_2} dt \left\{L^D+\frac{1}{2}\partial _t(P_0\vh_0)
\right\},~~~~~~~~
L^{D}=L+\int d^3x(P_{\cal N}\dot {\cal N} + P_{N^i}\dot N^i
- \lambda^0 P_{\cal N} - \lambda^i P_{N^i}).
\ee
We can define extended Dirac Hamiltonian as
\be \label{HED}
H^{D} = N_0\left[-\frac{P_0^2}{4V_0}+ I^{-1} H(\vh_0)
\right]+ \int d^3 x(\lambda^0 P_{\cal N} + \lambda^i P_{N^i}).
\ee
The equations obtained from variation of $W^D$ with respect to
Lagrange multipliers are called first class primary constraints
\be \label{primary}
P_{\cal N} =0,~~~~~~~~~~~~~~~~ P_{N^i}=0.
\ee
The condition of conservation of these constraints in time leads to
the first class secondary constraints
\be \label{secondary}
\left\{H^{D},P_{\cal N}\right\}=
{\cal H}-\frac{\int d^3x {\cal NH}}{V_0{\cal N}^2}=0,
~~~~~~~~~~
\left\{H^{D},P_{N^i}\right\}={\cal P}_i =0.
\ee
For completeness of the system we have to include set of secondary
constraints. According Dirac we choose them in the form
\be \label{gauge1}
{\cal N}(t, \vec x) =1; ~~~~~~~~~~~ N^i(t, \vec x)  = 0;
\ee

\be \label{gauge2}
\pi(t, \vec x) =0; ~~~~~~~~~~~ \chi^j:=\partial_i(q^{-1/3}q^{ij})=0.
\ee
The equations of motion obtained for our system are

\be \label{eqED}
\frac{dF}{dT}=\frac{\partial H(\vh_0) }{\partial P_F },
~~~~~~~~
-\frac{dP_F}{dT}=\frac{\partial H(\vh_0) }{\partial F }~,
\ee
where $H(\vh_0)$ is given by the equation (\ref{fpt1}),
and we introduced reparametrization-invariant geometric time $T$
\be \label{proptime}
N_0 dt \buildrel{\rm def}\over= dT~.
\ee

\subsection{Global constraints and equations of motion.}

The physical meaning of the geometric time $T$, the dynamic variable
$\vh_0$ and its momentum is given by
the explicit resolving of the zero-Fourier harmonic of the energy
constraint
\be \label{grconstr}
\frac{\delta W^E}{\delta N_0(t)}=
-\frac{P_0^2}{4V_0}+H(\vh_0) =0.
\ee
This constraint has two solutions for the global momentum $P_0$:
\be\label{glop}
(P_0)_{\pm}=\pm 2 \sqrt{V_0 H(\vh_0) }\equiv H^*_{\pm}.
\ee
The equation of motion for this global momentum $P_0$ in
gauge (\ref{gauge1}) takes the form
\be \label{nono}
\frac{\delta W^E}{\delta P_0}=0\,
\Rightarrow\,
\left(\frac{d\vh}{dT}\right)_{\pm}
=\frac{(P_0)_{\pm}}{2V}
=\pm\sqrt{\rho({\vh_0})};~~~\rho(\vh_0) =\frac{\int d^3x  {\cal H}}{ V_0}=
\frac{H(\vh_0) }{V_0}~.
\ee
The integral form of the last equation is
\be
\label{70}
T( {\vh_1},{\vh_2})=\int\limits_{\vh_1}^{\vh_2}d\vh
{\rho}^{-1/2}(\vh).
\ee
Equation obtained by varying the action with respect to $\vh_0$ follows
independently from the set of other constraints and equations of motion.

Equations (\ref{nono}), (\ref{70}) in the homogeneous approximation of GR
are the basis of observational cosmology where the geometric time is the
conformal time connected with the world time $T_f$ of the Friedmann cosmology
by the relation
\be \label{fried}
dT_f=\frac{\vh_0(T)}{\mu}dT~,
\ee
and the dependence of scale factor (dynamic evolution parameter $\vh_0$)
on the geometric time $T$ is treated as the evolution of the universe.
In particular, equation (\ref{nono}) gives the relation between the
present-day value of
the dynamic evolution parameter $\vh_0(T_0)$ and cosmological observations,
i.e., the density of matter $\rho$ and the Hubble parameter
\be \label{mu}
{\cal H}_{hub}^e
=\frac{\mu\vh_0'}{\vh_0^2}=
\frac{\mu\sqrt{\rho}}{\vh_0^2}~~~~\Rightarrow ~~~~
\vh_0(T_0)=\left(\frac{\mu\sqrt{\rho}}{{\cal H}_{hub}}\right)^{1/2}:=
\mu\Omega_{0}^{1/4} ~~~~~~~~(~0.6< (\Omega_0^{1/4})_{exp}<1.2~).
\ee
The dynamic evolution parameter as the cosmic scale factor and a value of its
conjugate momentum (i.e., a value of the dynamic Hamiltonian) as the density
of matter (see equations (\ref{nono}), (\ref{70}))
are objects of measurement in observational astrophysics and cosmology
and numerous discussions about the Hubble parameter, dark matter, and hidden
mass.

Our aim is to find the equivalent unconstrained Hamiltonian system
that describes evolution of the field world space $(F, \vh_0)$
in terms of the geometric time $T$.

\subsection{Equivalent Unconstrained Systems}

Assume that we can solve the constraint equations and pass to the reduced
space of independent variables ($F^*,P_F^*$).
The explicit solution of the local and global constraints has two
analytic branches with positive and negative values for scale factor
momentum $P_0$ (\ref{glop}). Therefore, inserting solutions of all
constraints into the action we get two branches of the equivalent
Dynamic Unconstrained System (DUS)
\be \label{DUS}
W_{\pm}^{DUS}[F|\vh_0]
=\int\limits_{\vh_1}^{\vh_2} d{\vh_0}\left\{  \left[\int d^3 x
\sum_{F^*} P_F^* \frac{\partial F^*}{\partial \vh_0}\right] - H^*_{\pm}
+ \frac{1}{2} \partial_{\vh_0}(\vh_0 H^*_{\pm})\right\}~,
\ee
where $\vh_0$ plays the role of evolution parameter and $H^*_{\pm}$ (defined
by equation (\ref{glop}) plays the role
of the evolution Hamiltonian in the reduced phase space of independent
physical variables $(F^*, P_F^*)$ with equations of motion
\be \label{EDUS}
\frac{dF^*}{d\vh_0}=\frac{\partial H^*_\pm}{\partial P_F^* },
~~~~~~~~
-\frac{dP_F^*}{d\vh_0}=\frac{\partial H^*_\pm}{\partial F^* }~.
\ee
The evolution of the field world space variables $(F^*,\vh_0)$ with respect
to  the geometric time $T$ is not contained in DUS (\ref{DUS}).
This geometric time evolution is described by
supplementary equation for nonphysical momentum $P_0$ (\ref{nono}) that
follows from the initial extended system.

To get an equivalent unconstrained system in terms of the geometric time
(we call it the Geometric Unconstrained System (GUS)), we need the
Levi-Civita canonical transformation (LC) \cite{lc,sh,gkp} of the field world
phase space
\be \label{L-C}
(F^*,P^*_F|\vh_0,P_0)\Rightarrow (F_G^*,P_G^*|Q_0,\Pi_0)
\ee
which converts the energy constraint (\ref{grconstr}) into the new momentum
$\Pi_0$ (see the similar transformations for a relativistic
particle in Appendix A).

In terms of geometrical variables the action takes the form
\be \label{L-CW}
W^G=\int\limits_{t_1}^{t_2}dt\left\{\left[\int d^3x\sum_{F_G^*}{P_G^*}
\dot F_G^*\right]- \Pi_0\dot Q_0+N_0\Pi_0+\frac{d}{dt}S^{LC}\right\}
\ee
where $S^{LC}$ is generating function of LC transformations.
Then the energy constraint and the supplementary equation for the new
 momentum take trivial form
\be \label{trivial}\Pi_0=0~;~~~~
\frac{\delta W}{\delta \Pi_0}=0 ~~~~\Rightarrow ~~~~\frac{dQ_0}{dt}=
N_0 ~~~~ \Rightarrow ~~~~{dQ_0}={dT}.
\ee
Equations of motion are also trivial
\be \label{GUSE}
\frac{dP^*_G}{dT}=0,  ~~~~~~~~\frac{dF^*_G}{dT}=0,
\ee
and their solutions are given by the initial data
\be \label{initialG}
P^*_G=  {P^*_G}^0,~~~~~~~~F^*_G=  {F^*_G}^0.
\ee
Substituting solutions of (\ref{trivial}) and (\ref{GUSE}) into
the inverted Levi-Civita transformations
\be \label{L-C-reverse}
F^*=F^*(Q_0, \Pi_0|F^*_G, P^*_G), ~~~~~  \vh_0=\vh_0(Q_0, \Pi_0|F^*_G, P^*_G)
\ee
and similar for momenta, we get formal solutions of (\ref{EDUS})
and (\ref{70})
\be \label{sol-DUS}
F^*=F^*(T, 0|{F^*_G}^0, {P^*_G}^0), ~~~~~
P_F^*=P_F^*(T, 0|{F^*_G}^0, {P^*_G}^0),
~~~~~ \vh_0=\vh_0(T, 0|{F^*_G}^0, {P^*_G}^0).
\ee
We see that the geometric time evolution of the dynamic variables
is absent in DUS.
The geometric time evolution of the dynamic variables can be described
in the form of the  LC (inverted) canonical transformation
of GUS into DUS (\ref{L-C-reverse}), (\ref{sol-DUS}).

To obtain the geometric time Hamiltonian evolution, it is sufficient
to use the weak form of Levi-Civita-type transformations to GUS
$(F^*,P^*_F) ~~ \Rightarrow ~~ (\tilde F, \tilde P)$
with a new constraint
\be \label{weak-constraint}
\tilde\Pi_0-\tilde H(\tilde Q_0, \tilde F, \tilde P)=0.
\ee
We get the constraint shell action
\be \label{weak-action}
\tilde W^{GUS}=\int dT \left\{\left[\int d^3x\sum_{\tilde F}\tilde P
\frac{d\tilde F}{dT}\right] - \tilde H(T,\tilde F, \tilde P)
\right\},
\ee
that allows us to choose the initial cosmological data with respect to the
geometric time.

The considered invariant reduction
reveals the difference of reparametrization-invariant theory
from the gauge-invariant theory: in gauge-invariant theory
the superfluous (longitudinal) variables are completely excluded
from the reduced system; whereas, in reparametrization-invariant theory
the superfluous (longitudinal) variables leave  the sector of
the Dirac observables (i.e., the phase space ($F^*,P_F^*$))
but not the sector of measurable quantities: superfluous (longitudinal)
variables become the dynamic evolution parameter and dynamic Hamiltonian
of the reduced theory.

\subsection{Quantization and the arrow of the time}

In quantum theory of GR (like in quantum theories of a particle
 considered in Appendix A), we get two Schr\"odinger equations
\be \label{sch1}
 i\frac{d}{d \vh_0}\Psi^{\pm}(F|\vh_0,\vh_1)=
H^*_{\pm}(\vh_0)\Psi^{\pm}(F|\vh_0,\vh_1)
\ee
with positive and negative eigenvalues of $P_0$
and normalizable wave functions with the spectral series
over quantum numbers $Q$
\be \label{psi+}
\Psi^+(F|\vh_0,\vh_1)=
\sum\limits_{Q }^{ }
A^+_Q <F|Q><Q|\vh_0,\vh_1>\theta (\vh_0-\vh_1)
\ee
\be \label{psi-}
\Psi^-(F|\vh_0,\vh_1)=
\sum\limits_{Q }^{ }
A_Q^-<F|Q>^*<Q|\vh_0,\vh_1>^*\theta (\vh_1-\vh_0)~,
\ee
where $<F|Q>$ is the eigenfunction of the reduced energy~(\ref{glop})
\be \label{eig}
H^*_{\pm}(\vh_0)<F|Q>=\pm E(Q,\vh_0)<F|Q>
\ee
\be \label{fq}
<Q|\vh_0,\vh_1>=\exp[-i\int\limits_{\vh_1 }^{\vh_0 }d\vh E(Q,\vh)],~~~~~~
 <Q|\vh_0,\vh_1>^*=\exp[i\int\limits_{\vh_1 }^{\vh_0 }d\vh E(Q,\vh)]~.
\ee
The coefficient $A^+_Q$, in "secondary" quantization, can be treated
as the operator of creation of a universe with positive energy;
and the coefficient $A^-_Q$, as the operator of
annihilation of a universe also with positive energy.
The "secondary" quantization means $[A_Q^-,A_{Q'}^+]=\delta_{Q,Q'}$.
The physical states of a quantum universe  are formed by the action of these
operators on the vacuum $<0|$, $|0>$ in the form of out-state
($|Q>=A_Q^+|0>$) with positive "frequencies" and in-state ($<Q|=<0|A_Q^-$)
with negative "frequencies".
This treatment means that positive frequencies propagate forward
(${\vh_0}>{\vh}_1$);
and negative frequencies, backward (${\vh}_1>{\vh}_0$), so that
the negative values of energy are excluded from the spectrum
to provide the stability of the quantum system in quantum theory of GR.
In other words, instead of  changing the sign of energy,
we change that of the dynamic evolution parameter, which leads to
the causal Green function
\be \label{causgr}
G_c(F_1,\vh_1|F_2,\vh_2)=G_+(F_1,\vh_1|F_2,\vh_2)\theta(\vh_2-\vh_1) +
G_-(F_1,\vh_1|F_2,\vh_2)\theta(\vh_1-\vh_2)
\ee
where $G_+(F_1,\vh_1|F_2,\vh_2)=G_-(F_2,\vh_2|F_1,\vh_1)$
is the "commutative" Green function
\be \label{FIpgr}
G_{+}(F_2,\vh_2|F_1,\vh_1)= <0|\Psi^-(F_2|\vh_2,\vh_1)
\Psi^+(F_1|\vh_1,\vh_1)|0>
\ee
For this causal convention, the geometric time~(\ref{70})
is always positive in accordance with the equations of motion~(\ref{nono})
\be \label{atgr}
\left(\frac{d T}{d \vh_0}\right)_{\pm}=\pm \sqrt{\rho}~~
\Rightarrow~~T_{\pm}({\vh_1},{\vh_0})=
\pm\int\limits_{\vh_1}^{\vh_0}d\vh{\rho}^{-1/2}(\vh) \geq 0~.
\ee
Thus, the causal structure of the field world space immediatly
leads to the arrow of the geometric time~(\ref{atgr})
and the beginning of evolution of a
universe with respect to the geometric time $T=0$.

As it was shown in~\cite{ps1}, the way to obtain conserved
integrals of motion in classical theory and quantum numbers $Q$
in quantum theory is the Levi-Civita-type canonical transformation
of the field world space $(F,\vh_0)$ to a geometric set of variables
($V,Q_0$) with the condition that the geometric evolution parameter $Q_0$
coincides with the geometric time $dT=dQ_0$ (see Fig. 3).

\section{Reparametrization invariant path integral}

Following Faddeev-Popov procedure we can write down the path
integral for local fields of our theory using constraints and gauge
conditions (\ref{primary}-\ref{gauge2}):
\be \label{idfp-local}
Z_{\rm local}(F_1,F_2|P_0,\vh_0,N_0)=\int\limits_{F_1}^{F_2}
D(F,P_f) \Delta_s \bar \Delta_t
\exp\left\{ i\bar W\right\},
\ee
where
\be \label{prod}
D(F,P_f)=
\prod\limits_{t,x}
\left(\prod\limits_{i<k} \frac{dq^{ik}d\pi_{ik}}{2\pi}
\prod\limits_{f }^{ }\frac{df dp_f}{2\pi} \right)
\ee
are functional differentials for the metric fields ($\pi, q$) and the
matter fields ($p_f, f$),
\be \label{fps}
\Delta_s=
\prod\limits_{t,x,i}
\delta({\cal P}_i)) \delta(\chi^j)det\{{\cal P}_i,\chi^j\} ,
\ee
\be \label{fpt}
\bar\Delta_t=
\prod\limits_{t,x}
\delta({\cal H}(\mu))\delta(\pi) det\{ {\cal H}(\vh_0)-\rho,\pi\},~~~~~~~~
\left(\rho=\frac{\int d^3x H(\vh_0)}{V_0}\right)
\ee
are the F-P determinants, and
\be \label{Egrc1}
\bar W=\int\limits_{t_1}^{t_2} dt \left\{\int\limits_{V_0} d^3x
\left(\sum\limits_{F }^{ }P_F\dot F \right)- P_0 \dot\vh_0
-N_0\left[-\frac{P_0^2}{4V_0}+  H(\vh_0)
\right]+\frac{1}{2}\partial _t(P_0\vh_0)
\right\}
\ee
is extended action of considered theory.

By analogy with SR considered in Appendix A we define a commutative
Green function as an integral over global fields $(P_0,\vh_0)$
and the average
over reparametrization group parameter $N_0$
\be \label{idfp}
G_+(F_1,\vh_1|F_2,\vh_2)=\int\limits_{\vh_1 }^{\vh_2 }
\prod\limits_{t }^{ }\left(\frac{d\vh_0 dP_0d\tilde N_0}{2\pi} \right)
Z_{\rm local}(F_1,F_2|P_0,\vh_0,N_0),
\ee
where
\be \label{average}
\tilde N={N}/{2\pi\delta(0)},~~~~~~ \delta(0)=\int dN_0.
\ee
The causal Green function in the world field space ($F,\vh_0$)
is defined as the sum
\be \label{cgfu}
G_c(F_1,\vh_1|F_2,\vh_2)=G_+(F_1,\vh_1|F_2,\vh_2)\theta(\vh_1-\vh_2)
+G_+(F_2,\vh_1|F_2,\vh_1)\theta(\vh_2-\vh_1).
\ee
This function will be considered as generating functional for
the unitary $S$-matrix elements ~\cite{bww}
\be \label{tr1}
S[\vh_1,\vh_2]=<{\mbox out}~(\vh_2)|
T_{\vh} \exp\left\{-i\int\limits_{\vh_1}^{\vh_2}d\vh (H^*_I)\right\}
|(\vh_1)~{\mbox in}>,
\ee
where $T_{\vh}$ is a symbol of ordering with respect to parameter $\vh_0$,
and $<{\rm out}~(\vh_2)|,~ |(\vh)~ {\rm in }>$ are states of
quantum Univers in the lowest order of the Dirac perturbation theory
($ {\cal N}=1;~~N^k=0;~q^{ij}=\delta_{ij}+h^T_{ij} $),
 $H^*_I$ is the interaction Hamiltonian
\be \label{int}
H^*_I = H^*-H^*_0,~~~H^*=2\sqrt{V_0 H(\vh)}, ~~~~~
H_0^*=2\sqrt{V_0 H_0(\vh)}~,
\ee
$H_0$ is a sum of the Hamiltonians of "free" fields
(gravitons, photons,  massive vectors, and spinors) where  all masses
(including the Planck mass) are replaced by the
dynamic evolution parameter $\vh_0$~\cite{ps1}. For example for gravitons
the "free" hamiltonian takes the form:
\be \label{hom}
H_0(\vh_0)=\int d^3x\left(\frac{6(\pi^T_{(h)})^2}{\vh_0^2}+\frac{\vh_0^2}{24}
(\ik_ih^T)^2\right);~~(h^T_{ii}=0;~~\ik_jh^T_{ji}=0).
\ee
In order to reproduce Faddeev-Popov integral for general relativity
in infinite space-time \cite{fp2}, one should fix the
dynamic evolution parameter at its present-day value $\vh_0=\mu$~(\ref{mu}),
remove all the zero-mode dynamics $P_0=\dot \vh_0=0, N_0=1$, and neglect
the surface Newton term in the Hamiltonian. We get
\be \label{zfp}
Z^{FP}(F_1,F_2)=Z_{\rm local}(F_1,F_2|P_0=0,{\vh_0}_{exp}=\mu, N_0=1),
\ee
or
\be \label{zfp1}
Z^{FP}(F_1,F_2)=
\int\limits_{F_1 }^{F_2 }
D(F,P_f) \Delta_s\Delta_t
\exp\left\{ i W_{fp}\right\},
\ee
where
\be \label{fp1}
W_{fp}=\int\limits_{-\infty}^{+\infty} dt \int d^3x
\left(\sum\limits_{F }^{ }P_F\dot F -  {\cal H}_{fp}(\mu)
\right),~~~~~{\cal H}_{fp}(\mu)={\cal H}(\mu)-
\frac{\mu^2}{6} \partial_i \partial_jq^{ij}~ ,
\ee
and
\be \label{fpt2}
\Delta_t=
\prod\limits_{t,x}
\delta({\cal H}(\mu))\delta(\pi) det\{ {\cal H}(\mu),\pi\}.
\ee
The F-P integral~(\ref{zfp1}) is considered as the generating functional for
unitary perturbation theory in terms of S-matrix elements
\be \label{tr3}
S[-\infty|+\infty]=<{\mbox out}|
T \exp\left\{-i\int\limits_{-\infty}^{+\infty} dt
H_I(\mu)\right\}
|{\mbox in}>.
\ee
Strictly speaking,
the approximation (\ref{zfp}) is not a correct procedure, as it breaks the
reparame\-tri\-za\-tion-invariance.
The region of validity of FP integral (\ref{zfp1}) is discussed in
next sections.

\section{Evolution of "Free" Quantum Universe}

\subsection{Dynamic unconstrained system}

Possible states of a free quantum universe in $S$-matrix (\ref{tr1})
are determined by
the lowest order of  the Dirac perturbation theory
given by the well-known system
of "free" conformal fields~(\ref{lich}),~(\ref{hom})
in a finite space-time volume~\cite{gmm,ps1}
\be \label{e0}
W_0^E= \int\limits_{t_1}^{t_2}dt
\left( \left[\int d^3x \sum\limits_{F}P_F\dot F \right]
-P_0\dot \vh_0 - N_{0}[-\frac{P_0^2}{4V}+H_0(\vh_0)]
+\frac{1}{2}\ik_0(P_0\vh_0)\right),
\ee
where $H_0$ is a sum of the Hamiltonians of "free" fields
(gravitons~(\ref{hom}), photons,  massive vectors, and spinors)
where all masses (including the Planck mass) are replaced by the
dynamic evolution parameter $\vh_0$~\cite{ps1}.
The classical equations for the action~(\ref{e0})
\be \label{eqED0}
\frac{dF}{dT}=\frac{\partial H_0}{\partial P_F },
~~~~~~~~
-\frac{dP_F}{dT}=\frac{\partial H_0}{\partial F },
~~~~~~~~ P_0=\pm 2\sqrt{V_0 H_0}:=H_0^*
\ee
contain two invariant times: the geometric $T$  and the dynamic
$\vh_0^{\pm}$  connected by
the geometro-dynamic (back-reaction) equation
\be \label{ea}
\frac{d\vh_0^{\pm}}{dT} = \pm\sqrt{\rho_0(\vh_0^{\pm})},
~~~~~~\left(\rho_0=\frac{H_0}{V_0}\right).
\ee
Solving the energy constraint we get the action for dynamic system
\be \label{e0d}
W_0^E(constraint)= W_0^D=\int\limits_{\vh(t_1)}^{\vh(t_2)}d\vh
\left( \left[\int d^3x \sum\limits_{F}P_F\partial_{\vh}  F \right]
- H^*_{0\pm} + \frac{1}{2}\partial_{\vh}(\vh H^*_{0\pm} ) \right),
\ee
that has two branches for a universe with
a positive energy $(P_0 > 0)$, and a universe with a negative energy
$(P_0 < 0)$.
We interpret the branch with negative energy as  an "anti-universe" which
propagates backward $(\vh <0 )$ with
positive energy to provide the stability of a quantum system.

The content of matter in the universe
is described by the number of particles $N_{F,k}$ and their energy
$\omega_F(\vh_0,k)$ (which depends on the dynamic evolution parameter $\vh_0$ and
quantum numbers $k$, momenta, spins, etc.).
Detected particles are defined as the field variables $F=f$
\be \label{grep}
   f(x)=\sum\limits_{k }^{ }
\frac{C_f(\vh_0)\exp(ik_ix_i)}{V_0^{3/2}\sqrt{2\omega_f(\vh_0,k)}}
\left( a_f^+(-k)+ a_f^-(k)\right)
\ee
which diagonalize the operator of the density of matter
\be \label{dide}
\rho_0=\sum\limits_{f, k }^{ } \frac{\omega_{f}(\vh_0,k)}{V_0}\hat N_{f,k},
~~~~~~~~
\hat N_f(a)=\frac{1}{2}(a_f^+a_f^- + a_f^-a_f^+)~.
\ee
We restrict ourselves to gravitons (f=h)
$C_h(\vh_0) = \vh_0\sqrt{12},~\omega_h(\vh_0,k)=\sqrt{k^2}$
and massive vector particles (f=v)
$C_v(\vh_0)=1,~\omega_v(\vh_0,k) =\sqrt{k^2+y^2\vh_0^2}$,
where $y$ is the mass in terms of the Planck constant.

\subsection{Geometric unconstrained system}

The equations of motion~(\ref{ea}) in terms of $a^+,a^-$~\cite{ps1}
are not diagonal
\be \label{nond}
i\frac{d}{dT}\chi:=i\chi'_{a_f} =-\hat H_{a_f} \chi_{a_f},~~~~
\chi_{a_f}=\left(\begin{array}{c} a_f^+\\a_f \end{array}\right);
\;\;{\hat H}_{a_f}=\left|
\begin{array}{ccc}\omega_{a_f} &,&-i\Delta_f\\ \\ -i\Delta_f&,&-\omega_{a_f}\end{array}
\right|~,
\ee
where nondiagonal terms $\Delta_{f=h,v}$
are proportional to the Hubble parameter~(\ref{mu})~\cite{ps1}
\be \label{delt}
\Delta_{f=h}=\frac{\vh_0'}{\vh_0},~~~~~~~~~~~~~~
\Delta_{f=v}=-\frac{\omega'_v}{2\omega_v},
~~~~~~~~~~  \vh_0'=\sqrt{\rho_0}~.
\ee

The "geometric system"  $(b^+,b)$ is determined by the transformation to the
set of variables which
diagonalize equations of motion~(\ref{eqED0})
and determine a set of integrals of motion of
equations~(\ref{eqED0}) (as conserved numbers $\left\{Q\right\}$).

To obtain integrals of motion and to choose initial conditions for
the universe evolution we use the Bogoliubov
transformations \cite{b} and define "quasi-particles"
\be \label{17}
b^+=\cosh(r)e^{-i\theta}a^+-i\sinh(r)e^{i\theta}a,~~~~~~~~
\;\;\;\;b=\cosh(r)e^{i\theta}a+i\sinh(r)e^{-i\theta}a^+~,
\ee
or
$$
\chi_b=\left(\begin{array}{c} b^+\\b \end{array}\right)=
\hat O\chi_{a}~,
$$
which diagonalize the classical equations expressed in terms
of "particles" $(a^+,a)$, so that the "number of quasiparticles"  is
 conserved
\be \label{17a}
 \frac{d(b^+b)}{dt}=0,~~~~~~~~~~~
b=\exp(-i\int\limits_{0 }^{T }d\bar T \bar \omega_b(\bar T))b_0.
\ee
Functions $r$ and $\theta$ in ~(\ref{17}), and the quasiparticle
energy $\bar \omega_b$  in~(\ref{17a}) are
determined by the equation of diagonalization
\be \label{diago}
i\frac{d}{dT}\chi_b=[-i\hat O{}^{-1}\frac{d}{dT}\hat O -
\hat O{}^{-1}\hat H{}_a\hat O]\chi_b\equiv -
\left(
\begin{array}{rr} \bar \omega_b,&0\\ \\ 0,&-\bar \omega_b\end{array}\right)\chi_b
\ee
in the form obtained in~\cite{ps1}
\be \label{17b}
\bar \omega_{f b}=(\omega_f-\theta'_f)\cosh(2r_f)-
(\Delta_f\cos2\theta_f)\sinh(2r_f)~,~~~~~~~~~~~~~~~~~~~~
\ee
$$
0 = (\omega_f-\theta'_f)\sinh(2r_f)-(\Delta_f\cos2\theta_f)\cosh(2r_f),
~~~~~~~r'_f=-\Delta_f\sin2\theta_f~.
$$
Equations~(\ref{delt})--~(\ref{17b}) are closed by
the definition of "observable particles" in terms of quasiparticles
\be \label{obs}
\rho(\vh)=\frac{H_0}{V}=
\frac{\sum\limits_f\omega_f(\vh)\{a_f^+a_f\}}{V_0};~~~~~
\{a^+a\}=\{b_0^+b_0\}\cosh{2r}-\frac{i}{2}(b^{+2}-b^2)\sinh{2r}
\ee
with
\be \label{17c}
\bar \omega_{f b}=\sqrt{(\omega_f-\theta_f')^2+(r'_f)^2-\Delta_f^2},~~~~~
\theta_f'=-\frac{1}{2}\left(\frac{r'_f}{\Delta_f}\right)'
[1-\frac{(r'_f)^2}{\Delta_f^2}]^{-1/2},~~~
\cosh(2r_f)=\frac{\omega_f-\theta_f'}{\bar \omega_{f b}}~.
\ee

The constrained system in terms of geometric variables is described
by the action
\be \label{transparence1}
\tilde W^G=\int dt\left\{\sum_f \frac{i}{2}
\left(b\partial_t b^+-b^+\partial_t b\right)_f
-\tilde\Pi_0\dot Q_0 - N_0\left[-\tilde\Pi_0+\sum_f \omega^f_b(Q_0) N_f(b)\right]\right\}~,
\ee
where the new dynamic evolution parameter $Q_0$ coincides with
geometric time $T$ on the equations of motion
\be \label{dQ}
\frac{\delta \tilde W^E}{\delta\tilde\Pi_0}=0 ~~~~ \Rightarrow ~~~~\frac{dQ_0}{dt}
~~~~ \Rightarrow dQ_0=dT~.
\ee
Reduction of this system leads to the weak version of Geometric
Unconstrained System (\ref{weak-action})
\be \label{transparence2}
\tilde W^{GUS}=\int dT\left\{\sum_f \frac{i}{2}
\left(b\partial_T b^+-b^+\partial_T b\right)_f
-\sum_f \omega^f_b(T) N_f(b)\right\}~.
\ee
We choose the initial data appropriate for the dynamics described by GUS
(\ref{transparence2}).

\subsection{Quantization}

The initial data $b_0,b_0^+$ of quasiparticle variables (\ref{17a})
form the
set of quantum numbers in quantum theory.

Let us suppose that we manage to solve equations~(\ref{17a})--~(\ref{17c})
with respect to the geometric time $T$ in terms of conserved numbers
$b^+_0,b_0$. This means that
the wave function of a quantum universe can be represented
in the form of a series over the conserved quantum
numbers $Q=n_{f,k}=<Q|b_f^+b_f|Q>$ of the Bogoliubov states
\be \label{ewfg}
\Phi_Q(T)=
\prod\limits_{f,n_f }^{ }
\exp\left\{-i\int\limits_{0 }^{T }dT
n_f \bar \omega_b(T)\right\}\frac{(b_f^+)^{n_f}}{\sqrt{n_f!}}|0>_b~.
\ee
In this geometric system, we have an arrow of
the geometric time $T$ for the universe
\be \label{u1}
T_+(\vh_2,\vh_1)= \int\limits_{\vh_1 }^{\vh_2 } d\vh \rho(\vh)^{-1/2} > 0,~~
~~~~~\vh_2 > \vh_1~,
\ee
and for anti-universe
\be \label{u}
T_-(\vh_2,\vh_1)=-
 \int\limits_{\vh_1 }^{\vh_2 } d\vh \rho(\vh)^{-1/2}=
 \int\limits_{\vh_2 }^{\vh_1 } d\vh \rho(\vh)^{-1/2} > 0,~~
~~~~~\vh_1 > \vh_2.
\ee
The dynamic system~(\ref{e0d}) of particle variables $a^+,a$
is connected with
the geometric one by the Bogoliubov transformations. Using
these transformations we can find wave functions of a universe,
for $\vh_2>\vh_1$
and an anti-Universe, for $\vh_1>\vh_2$
\be \label{ewf}
\Psi_Q(T)=
A_Q^+\Phi_Q(T_+(\vh_2,\vh_1))\theta(\vh_2-\vh_1)+
A_Q^-\Phi_Q^*(T_-(\vh_2,\vh_1))\theta(\vh_1-\vh_2)~,
\ee
where the first term and the second one are positive ($P_0>0$)
and negative ($P_0<0$) frequency parts of the solutions with the
spectrum of quasiparticles $\bar \omega_b$,
$A_Q^{+}$ is the operator
of creation of a universe with a positive
"frequency" (which propagates in the positive direction of the dynamic
evolution parameter) and $A_Q^{-}$ is the operator
of  annihilation of a universe (or creation of an anti-universe) with
a negative "frequency" (which propagates in the negative direction of the
dynamic evolution parameter).

We can see that the creation of the universe in the field world space and the
creation of dynamic particles by the geometric vacuum $(b^+|0>=0)$
are two different effects.

The second effect disappears if we neglect gravitons and massive fields.
In this case, $d\rho /d\vh =0$, and one can represent a wave function of
Universe in the form of the spectral series over eigenvalues
$\rho_Q$ of the density $\rho$
\be \label{ewfr}
\Psi(f|\vh_2,\vh_1)=
\ee
$$
\sum\limits_{Q }^{ } \left[
\frac{A^+_Q}{\sqrt{2\rho_Q}}\exp\left\{-i (\vh_2-\vh_1)
\sum\limits_{ }^{ }\frac{\bar \omega_f n_f}{\sqrt{\rho_Q}}\right\}<f|Q>+
\frac{A^-_Q}{\sqrt{2\rho_Q}}\exp\left\{i (\vh_2-\vh_1)
\sum\limits_{ }^{ }\frac{\bar \omega_f n_f}{\sqrt{\rho_Q}}\right\}<f|Q>^*
\right],
$$
where $<f|Q>$ is a product of normalizable Hermite polynomials.

\subsection{Evolution of quantum universe}

The equations of diagonalization~(\ref{diago}) for the Bogoliubov
coefficients~(\ref{17}) and the quasiparticle energy $\bar \omega_b$
(\ref{17b}) play the role
of the equations of state of the field matter in the universe.
We can show that the choice of
initial conditions for the "Big Bang" in the form of the Bogoliubov
(squeezed) vacuum $b|0>_b=0$ reproduces all stages of the evolution
of the Friedmann-Robertson-Walker universe in their conformal versions:
anisotropic, inflation, radiation, and dust.

The squeezed vacuum (i.e., the vacuum of quasiparticles)
is the state of "nothing".
For small $\vh$ and a large Hubble parameter,
at the beginning of the universe, the state of  vacuum of quasiparticles
leads to the density of matter~\cite{ps1}
\be \label{early}
{}_b<\rho(a^+,~a)>_b = \rho_0 \frac{1}{2} \left(\frac{\vh^2(0)}{\vh^2(T)} +
\frac{\vh^2(T)}{\vh^2(0)}\right),~~~~~~~\theta=\frac{\pi}{4}
\ee
where $\vh(0)$ is the initial value, and $\rho_0$
is the density of the Casimir energy of vacuum of "quasiparticles".
The first term corresponds to the conformal version of the
rigid state equation (in accordance with the
classification of the standard cosmology) which describes
the Kasner anisotropic stage $T_{\pm}(\vh)\sim \pm\vh^2$
(considered on the quantum level by Misner~\cite{M}).
The second term of the squeezed vacuum density~(\ref{early})
(for an admissible positive branch) leads to
the stage with inflation of the dynamic evolution parameter $\vh$ with respect to the
geometric time $T$
$$
\vh(T)_{(+)}\simeq \vh(0)\exp[T\sqrt{2\rho_0}/\vh(0)].
$$
It is the stage of intensive creation of "measurable particles".
After the inflation,  the Hubble parameter goes to zero, and
 gravitons convert into photon-like oscillator excitations with
the conserved number of particles.

At the present-day stage, the Bogoliubov quasiparticles coincide
with particles, so that the measurable density of energy of matter
in the universe is a sum of relativistic energies of all particles
\be \label{acc}
\rho_0(\vh)=\frac{E}{V_0}=
\sum\limits_{n_f }^{ } \frac{n_f}{V_0} \sqrt{k_{fi}^2+y_f^2\vh^2(T)},
\ee
where $y_f$ is the mass of a particle in units of the Planck mass.
The case of massless particles ($y=0, \rho_0(\vh)= constant$) correspond to
the conformal version of radiation stage of the standard FRW-cosmology.
And the massive particles at rest
($k=0, \rho_0(\vh)=\rho_{\rm barions}\vh/\mu)$
corresponds to the conformal version of the dust universe of
the standard cosmology with the Hubble law
\be \label{dust}
\vh'=\pm \sqrt{\rho_0}~ \Rightarrow~~\vh_{\pm}(T)=
\left(\frac{\rho_{\rm barions}}{4\mu}\right) T^2;
~~~~~q=\frac{\vh^{\prime\prime}\vh}{{\vh^{\prime}}^2}=\frac{1}{2}.
\ee
The dynamic evolution parameter is expressed through the geometric time
of a quantum asymptotic state of the universe
$|out>$ and  conserved quantum numbers of this state:
energy $E_{out}$ and density $\rho_{0}=E_{out}/V_0$.

It is well-known that $E_{out}$ is a tremendous energy $(10^{79} GeV)$ in
comparison  with possible real and virtual deviations of the free Hamiltonian in
the laboratory  processes:
\be \label{lab1}
 \bar H_0 = E_{out} + \delta H_0 ,~~~
<{\mbox out}| \delta H_0|{\mbox in}> << E_{out}.
\ee
We have seen that
the dependence of the scale factor $\vh_0$ on the geometric time $T$
(or the "relation" of two classical
unconstrained systems: dynamic and geometric)
describes the "Big Bang" and evolution of a universe.

Therefore, from the point of view of unconstrained system
"Big Bang" is effect of evolution of the geometric interval with respect to
the dynamic evolution parameter which goes beyond  the scope of
Hamiltonian description of a single classical unconstrained system.

Reparametrization-invariant dynamic of General Relativity is covered
by Geometric and Dynamic Unconstrained Systems connected by the Levi-Civita
transformation of fields of matter into the
vacuum fields of initial data with respect to geometric time (see Fig. 3).

\section{QFT limit of  Quantum Gravity}

The simplest way to determine the QFT limit of  Quantum Gravity
and to find the region of
validity of the FP-integral(\ref{zfp1})
is to use the quantum field
version of the reparametrization-invariant
integral(\ref{idfp}) in the form of
S-matrix elements ~\cite{bww} (see (\ref{tr1}), (\ref{int})).
We consider the infinite volume limit of
the S-matrix element (\ref{int}) in terms
of the geometric time $T$ for the present-day
stage  $ T=T_0, \vh(T_0)=\mu$, and
$T(\vh_1)=T_0-\Delta T, T(\vh_2)=T_0+\Delta T=T_{out}$.
One can express this matrix element in terms of the time measured by
an observer of an out-state with a tremendous number of particles
in the Universe using equation ~(\ref{dust}) $d\vh=dT_{out}\sqrt{\rho_{out}}$
and approximation ~(\ref{lab1}) to neglect "back-reaction".
In the infinite volume limit, we get from (\ref{int})
\be \label{limit} d \vh_0 [ H^*_I ]=2 d \vh_0 \left(\sqrt{V_0(H_0+H_I)}-\sqrt{V_0 H_0}\right)=
dT_{out}[\hat F \bar H_I +  O(1/E_{out})]
\ee
where $ H_I$ is the interaction Hamiltonian in GR,
and
\be \label{form}
\hat F = \sqrt{\frac{E_{out}}{ H_0}}
= \sqrt{\frac{E_{out}}{E_{out} + \delta  H_0}}
\ee
is a multiplier which plays the role of a form factor for
physical processes  observed in  the "laboratory"  conditions
when the cosmic energy $E_{out}$ is much greater than the
deviation of the free energy
\be \label{lab}
\delta  H_0 =  H_0-E_{out};~~
\ee
due to creation and annihilation of
real and virtual particles in the laboratory experiments.

The measurable time of the laboratory experiments $T_2-T_1$ is much
smaller than the age of the universe $T_0$, but it is much greater
than the reverse "laboratory"  energy  $ \delta$, so that the
limit
$$
\int\limits_{T(\vh_1) }^{T(\vh_2) }dT_{\rm out}
\Rightarrow \int\limits_{-\infty }^{+\infty }dT_{\rm out}
$$
is valid.
If we neglect  the form factor ~(\ref{form}) that removes a set of
ultraviolet divergences, we get the matrix element~(\ref{tr3}) that
corresponds to the standard FP functional integral~(\ref{zfp1}) and
$S$-matrix element (\ref{tr3})
where the coordinate time is replaced by the geometric (conformal) time
$t \rightarrow T_{\rm out}$:
\be \label{tr4}
S[-\infty|+\infty]=<{\mbox out}|
T \exp\left\{-i\int\limits_{-\infty}^{+\infty} dT_{\rm out}\hat F
H_I(\mu)\right\}
|{\mbox in}> ~~~~~~~~ \left(\hat F=1\right).
\ee
Thus, the standard FP-integral (i.e., the Hamiltonian description of
the evolution of fields with respect to the geometric time) appears
as the nonrelativistic approximation of tremendous mass of a universe and
its very large life-time (see Fig. 4).
Now, it is evident that the conventional FP unitary S-matrix
are not valid for the description of the early universe given in the finite
spatial volume and the finite positive interval of geometrical time
when the early universe only begins to create matter.

On the other hand, we revealed that standard quantum field theory~(\ref{tr4})
is expressed in terms of the conformal-invariant Lichnerowicz variables
and coordinates including the conformal time ($T_{out}$) as the time
of evolution of these variables.
The reparametrization-invariant description of the "Big Bang"
evolution distinguishes conformal variables~(\ref{lich}) and coordinates.
The conformal invariance of the variables can testify to the conformal
invariance of the initial theory of gravity.
Such the theory can be a scalar version
of the Weyl conformal invariant theory~\cite{grg,plb,pr} (that
dynamically coincides with the Einstein General Relativity)
where these conformal variables, coordinates, geometric time $T$, and
the conformal Hubble parameter ${\cal H}^c_{hub}=\vh'/\vh$
can be considered as  measurable quantities.
In particular, to get "accelerating" universe with $q>0$ in the dust
stage (see (\ref{dust})), is enough to count that we measure the relative
Weyl time ($T_{\rm out}$) of quantum field theory (\ref{tr4}).

\section{ Conclusions}

The main result of the paper is the reparametrization-invariant
generating functional for the unitary and causal perturbation theory
in general relativity in a finite space-time.
We show that the classical cosmology of a universe
and the Faddeev-Popov-DeWitt functional correspond to  different orders of
decomposition of this functional over the inverse "mass" of a universe.

This result is based on the assertion that the
measurable time in any reparametrization-invariant system is not the
coordinate time, but the time-like dynamic variable of an extended phase
space of this system (of the type of the conformal scale factor). Accordingly, the measurable Hamiltonian is
a solution of the energy constraint
with respect to the conjugate momentum of this
dynamic evolution parameter ~\cite{grg,plb}.
This definition of the dynamic evolution parameter and Hamiltonian supposes the
reduction of an action for constructing an "equivalent" unconstrained
system.

The second assertion is that such an unconstrained dynamic system
cannot cover the physical content of a relativistic
reparametrization-invariant system.
This content can be covered by two
"equivalent" unconstrained systems -- the mentioned above dynamic system and
the geometric system constructed by the Levi-Civita type canonical
transformation so that
 a new dynamic evolution parameter coincides with the geometric time.

The "geometric" variables are
the Bogoliubov quasiparticles which diagonalize equations of motion
and give cosmological initial conditions.
The coefficients of the Bogoliubov transformations,
in the conventional QFT perturbation theory, determine the evolution density of matter. For the vacuum initial data this evolution reproduces all stages of
the standard FRW evolution of the universe in their conformal versions.

Consistent QFT limits of the generating functional
in classical gravitation and cosmology
and the conventional quantum field theory (in the form of the
Faddeev-Popov generating functional for a infinite space-time)
can be fulfilled for a tremendous value of the universe mass and
the universe life-time (see Fig. 4). The quantum field theory limit
distinguish the conformal treatment of general relativity
developed in \cite{grg,plb,pr}.

\bigskip\bigskip

{\bf Acknowledgments}

\medskip

We are happy to acknowledge interesting and critical
discussions with  Profs. B. M. Barbashov, A. Borowiec, A. M. Chervyakov, A.V. Efremov,
G.A. Gogilidze, V.G. Kadyshevsky, A.M. Khvedelidze, H. Kleinert, E.A. Kuraev,
D. Mladenov, V.V. Papoyan, V. I. Smirichinski, J. \'Sniatycki, I. V. Tyutin,
Yu. S. Vladimirov.
The work was supported
by the Committee for Scientific Research grant no. 603/P03/96 and by
the Infeld-Bogoliubov program.

\bigskip

\appendix
{\Large\bf APPENDIX A. Special Relativity}

\medskip

\renewcommand{\theequation}{A.\arabic{equation}}

\setcounter{equation}{0}

To answer the question: Why is the reparametrization-invariant
Hamiltonian reduction needed?, let us consider
 relativistic mechanics~\cite{grg} in the Hamiltonian form
\be \label{SR}
W[P,X|N|\tau_1,\tau_2]=
\int\limits_{\tau_1 }^{\tau_2 }d\tau [- P_{\mu}\dot X^{\mu}
- \frac{N}{2m}(-P_{\mu}^2+m^2) ]~.
\ee
This action is invariant with respect to reparametrizations of
the {\it coordinate evolution parameter} \be \label{coo1}
\tau \rightarrow \tau'=\tau'(\tau),~~~~~N'd\tau'=Nd\tau
\ee
given in the one-dimensional space with the invariant interval
\be \label{pros}
dT:=Nd\tau,~~~~~~~~T=\int\limits_{0 }^{\tau } d\bar \tau N(\bar \tau)~.
\ee
We called this invariant interval
the {\it geometric time}~\cite{grg} whereas
the dynamic variable $X_0$ (with a negative contribution
in the constraint) we called {\it dynamic evolution parameter}.

In terms of the geometric time~(\ref{pros})
the classical equations of the generalized Hamiltonian system~(\ref{SR})
takes the form
\be \label{cesr}
\frac{d X_{\mu}}{d T}=\frac{P_{\mu}}{m},~~~~~~~
\frac{d P_{\mu}}{d T}=0,~~~~~~~P_{\mu}^2-m^2=0.
\ee

The problem is
to obtain the equivalent unconstrained theories
directly in terms of the invariant times $X_0$ or $T$ with
the invariant Hamiltonians of
evolution with respect to these times.
The solution of this problem is called the dynamic (for $X_0$), or geometric
(for $T$) {\it reparametrization-invariant Hamiltonian reductions}.

The dynamic reduction of the extended system~(\ref{SR}) means
the substitution, into it, of
the explicit resolving of the energy constraint $(-P_{\mu}^2+m^2)=0$ with
respect to the momentum $P_0$ with a negative contribution
\be \label{po}
\frac{\delta W}{\delta N}=0~~\Rightarrow~ ~P_0=\pm\sqrt{m^2+P_i^2}.
\ee
In accordance with the two signs of the solution~(\ref{po}),
after the substitution of~(\ref{po}) into~(\ref{SR}), we have
two branches of the dynamic unconstrained system
\be \label{srd}
W(\mbox{constraint})= W^D_{\pm}[P_i,X_i|X_0(1),X_0(2)]=
\int\limits_{X_0(\tau_1)=X_0(1)}^{X_0(\tau_2)=X_0(2)} dX_0
\left[ P_{i}\frac{dX_i}{dX_0}\mp\sqrt{m^2+P_i^2} \right]~.
\ee
The role of the time of evolution,
in this action,
is played by the variable $X_0$ which abandons the Dirac sector of
"observables" $P_i, X_i$, but not the sector of
"measurable" quantities.
At the same time, its conjugate momentum $P_0$
converts into the corresponding Hamiltonian of evolution,
values of which are the energy of a particle.

This invariant reduction of the action gives the "equivalent"  unconstrained
system together with definition of the invariant evolution parameter
($X_0$)
corresponding to a non-zero Hamiltonian $P_0$.

Thus, we need the reparametrization-invariant Hamiltonian reduction to
determine the invariant evolution parameter and its invariant Hamiltonian for
reparametrization-invariant systems.

In quantum relativistic theory, we get two Schr\"odinger equations
\be \label{srw}
i\frac{d}{dX_0}\Psi_{(\pm)}(X|P)=\pm\sqrt{m^2+P_i^2}\Psi_{(\pm)}(X|P)~,
\ee
with positive and negative values of $P_0$
and normalized wave functions
\be \label{srnw}
\Psi_{\pm}(X|P)
=\frac{A_P^{\pm}\theta(\pm P_0)}{(2\pi)^{3/2}\sqrt{2P_0}}
\exp(-iP_{\mu}X^{\mu}),
~~~~~~~
\left([A_P^{-},A_{P'}^{+}]=\delta^3(P_i-P'_i)\right)~.
\ee
The coefficient $A_P^{+}$, in the secondary
quantization, is treated as the operator of creation
of a particle with positive energy;
and the coefficient $A_P^{-}$, as the operator of annihilation  of
a particle also with positive energy.
The physical states are formed by action of these operators on the vacuum
$<0|,|0>$ in the form of out-state (~$|P>=A_P^+|0>$~) with
positive frequencies
and in-state (~$<P|=<0|A_P^-$~) with
negative frequencies.
This treatment means that positive frequencies propagate forward
(${X_0}_2>{X_0}_1$);
and negative frequencies, backward (${X_0}_1>{X_0}_2$), so that
the negative values of energy are excluded from the spectrum
to provide the stability of the quantum system
in QFT~\cite{bww}. For this causal convention the geometric time~(\ref{pros})
is always positive in accordance with the equations of motion~(\ref{cesr})
\be \label{at}
\left(\frac{d T}{d X_0}\right)_{\pm}=\pm \frac{m}{\sqrt{P_i^2+m^2}}~~
\Rightarrow~~T({X_0}_2,{X_0}_1)=
\pm \frac{m}{\sqrt{P_i^2+m^2}}({X_0}_2-{X_0}_1) \geq 0
\ee
In other words, instead of  changing the sign of energy,
we change that of the dynamic evolution parameter, which leads to
the arrow of the geometric time~(\ref{at}) and to the causal Green function
\be \label{caus}
G^c(X)=G_+(X)\theta(X_0)+G_-(X)\theta(-X_0)=
i\int\limits_{ }^{ }\frac{d^4P}{(2\pi)^4}
\exp(-iPX)\frac{1}{P^2-m^2-i\epsilon},
\ee
where $G_+(X)=G_-(-X)$ is the "commutative" Green function~\cite{bww}
\be \label{FIp}
G_{+}(X)=\int\limits_{ }^{ }\frac{d^4P}{(2\pi)^3}
\exp(-iPX)\delta(P^2-m^2)\theta(P_0)=
\frac{1}{2\pi}\int\limits_{ }^{ }d^3Pd^3P'
<0|\Psi_{-}(X|P)\Psi_{+}(0|P')|0>~.
\ee

To obtain the reparametrization-invariant form of the functional integral
adequate to the considered gauge-less reduction~(\ref{srd}) and the
causal Green function~(\ref{caus}), we use
the version of composition law for the commutative Green function
with the integration over the whole measurable sector $X_{1 \mu}$
\be \label{Dl}
G_+(X-X_0)=\int\limits_{ }^{ } G_+(X-X_1)\bar G_+(X_1-X_0)dX_1~~~~~~~~~
(~\bar G_+=\frac{G_+}{2\pi\delta(0)}~)~,
\ee
where $\delta(0)=\int dN$ is the infinite volume of the group
of reparametrizations of the coordinate $\tau$.
Using the composition law $n$-times, we got
the multiple integral
\be \label{firp}
G_+(X-X_0)=\int\limits_{ }^{ }G_+(X-X_1)
\prod\limits_{k=1}^{n}\bar G_+(X_k-X_{k+1})dX_k~,~~~~~~~
(~X_{n+1}=X_0~)~.
\ee
The continual limit of the multiple integral with the integral representation
for $\delta$-function
$$
\delta(P^2-m^2)=\frac{1}{2\pi}\int\limits_{ }^{ } d N \exp[i N (P^2-m^2)]
$$
can be defined as the path integral
in the form of the average over the group of reparametrizations
\be \label{srfi}
G_+(X)=\int\limits_{X(\tau_1)=0 }^{X(\tau_2)=X }
\frac{dN(\tau_2)d^4P(\tau_2)}{(2\pi)^3}
\prod\limits_{\tau_1 \leq \tau < \tau_2}^{ }\left\{ d\bar N(\tau)
\prod\limits_{\mu}\left( \frac{dP_{\mu}(\tau)dX_{\mu}(\tau)}{2\pi}\right)
\right\}
\exp(i W[P,X|N|\tau_1,\tau_2]),
\ee
where $\bar N={N}/{2\pi\delta(0)}$,
and $W$ is the initial extended action~(\ref{SR}).

This functional integral has the form of
the average over the group of reparametrization
of the integral over the sector of "measurable" variables $P_{\mu}, X_{\mu}$.

The Hamiltonian unconstrained system in terms of the geometric time $T$
can be obtained by the canonical Levi-Civita - type
transformation~\cite{lc,sh,gkp}
\be
(P_{\mu}, X_{\mu}) \Rightarrow\, (\Pi_{\mu}, Q_{\mu})
\ee
to the variables ($\Pi_{\mu},Q_{\mu}$) for which one of equations
identifies $Q_0$ with the geometric time $T$.
This transformation~\cite{lc}
converts the constraint into a new momentum
\be \label{levi}
\Pi_0= \frac{1}{2m}[P_{0}^2 - P_i^2] ,~~~~~~\Pi_i=P_i,~~~~
 Q_0=X_0\frac{m}{P_0} ,~~~~~Q_i=X_i-X_0\frac{P_i}{P_0}
\ee
and has the inverted form
\be \label{ivel}
P_0=\pm \sqrt{2m\Pi_{0}+\Pi_i^2},~~~~P_i=\Pi_i,~~~~
X_0=\pm Q_0\frac{\sqrt{2m\Pi_{0}+\Pi_i^2}}{m},
~~~~~X_i=Q_i+Q_0\frac{\Pi_i}{m}.
\ee
After transformation~(\ref{levi}) the action~(\ref{SR}) takes the form
\be \label{SRlc}
 W=\int\limits_{\tau_1}^{\tau_2} d\tau
\left[
- \Pi_{\mu}\dot Q^{\mu}- N(-\Pi_0+  \frac{m}{2} )-\frac{d}{d\tau}S^{lc}
\right],~~~~S^{lc}=(Q_0 \Pi_0).
\ee
The invariant reduction is the resolving of the constraint $\Pi_0={m}/{2}$
which determines a new Hamiltonian
of evolution with respect to the new dynamic evolution parameter $Q_0$, whereas the equation
of motion for this momentum $\Pi_0$
identifies  the dynamic evolution parameter $Q_0$ with the geometric time $T$ ($Q_0=T$).
The substitution of these solutions into the action~(\ref{SRlc})
leads to the reduced action of a geometric unconstrained system
\be \label{geo}
W(\mbox{constraint})= W^G[\Pi,Q_i|T_1,T_2]=\int\limits_{T_1}^{T_2} dT
\left( \Pi_{i}\frac{dQ_i}{dT}- \frac{m}{2} - \frac{d}{dT} (S^{lc})  \right)
~~~~~(S^{lc}=Q_0\frac{m}{2}),
\ee
where variables $\Pi_i,Q_i$ are cyclic ones and have the meaning
of initial conditions in the comoving frame
\be
 \frac{\delta W}{\delta \Pi_i}=
\frac{dQ_i}{d\tau}=0
\Rightarrow\,
Q_i=Q_i^{(0)},~~~~
\frac{\delta W}{\delta Q_i}=
\frac{d\Pi_i}{d\tau}=0
\Rightarrow\,
\Pi_i=\Pi_i^{(0)}.
\ee
The substitution of all geometric solutions
\be
Q_0=T,~~\Pi_0=\frac{m}{2},~~\Pi_i=\Pi_i^{(0)}=P_i,~~Q_i=Q_i^{(0)}
\ee
into the inverted Levi-Civita transformation~(\ref{ivel}) leads to the
conventional  relativistic solution for the dynamical system
\be \label{line}
P_0=\pm \sqrt{m^2+P_i^2},
~~~~P_i=\Pi_i^{(0)},~~~~
X_0(T)= T\frac{P_0}{m},~~~~
X_i(T)=X_i^{(0)}+T\frac{P_i}{m}.
\ee
The Schr\"odinger equation for the wave function
\be \label{geom}
\frac{d}{idT}\Psi^{lc}(T,Q_i|\Pi_i)= \frac{m}{2}\Psi^{lc}(T,Q_i|\Pi_i),~~~~~~
\Psi^{lc}(T,Q_i|\Pi_i)=  \exp(-iT\frac{m}{2})
\exp(i\Pi_i^{(0)}Q_i)
\ee
contains only one eigenvalue $m/2$  degenerated with
respect to the cyclic momentum $\Pi_i$.
We see that there are differences between the dynamic and geometric
descriptions.
The dynamic evolution parameter is given in the whole region $-\infty < X_0 < +\infty$,
whereas the geometric one is only positive $0<T< +\infty$, as it
follows from the properties of the causal Green function~(\ref{caus})
after the Levi-Civita transformation~(\ref{levi})
$$
G^c(Q_{\mu})=\int\limits_{-\infty }^{+\infty }d^4\Pi_{\mu}
\frac{\exp(iQ^{\mu}\Pi_{\mu})}{2m(\Pi_0-m/2-i\epsilon/2m)}=
\frac{\delta^3(Q)}{2m}\theta(T),~~~~~~T=Q_0~.
$$
Two solutions of the constraint (a particle and antiparticle) in the dynamic
system correspond to a single solution in the geometric system.

Thus, the reparametrization-invariant content of the equations of motion of a
relativistic particle in terms of the geometric time is covered
by two "equivalent" unconstrained systems: the dynamic and geometric. In both
the systems, the invariant times are not {\sl the coordinate evolution
parameter}, but variables with the negative contribution into the energy constraint.
The Hamiltonian description of a relativistic particle in terms of the geometric
time can be achieved by the Levi-Civita-type canonical transformation, so that
the energy constraint converts into a new momentum.
Whereas, the dynamic unconstrained system is suit for the secondary quantization
and the derivation of the causal Green function that determine the arrow of the geometric time.

\vspace{1cm}

{\bf FIGURE CAPTIONS}

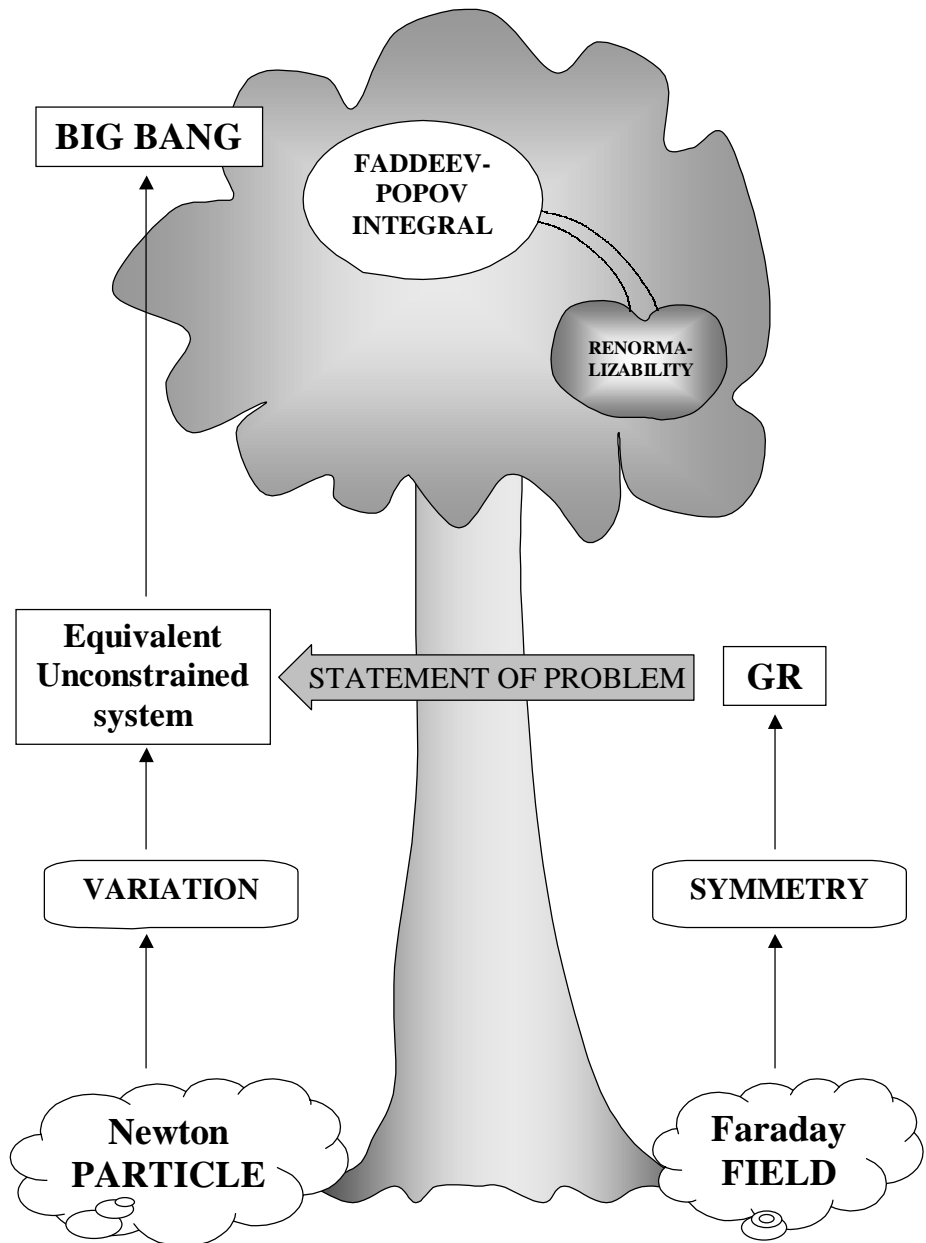
Fig. 1. The tree ot modern theoretical physics grew from two different
roots ("particle" and "field") which gave the VARIATIONAL method and
SYMMETRY principles for formulating  modern physical theories as
constraned systems. To obtain unambigious physical results, one should
construct Equivalent Unconstrained Systems compatible with
the symplest variational method. It is just the problem discussed
in the present paper.

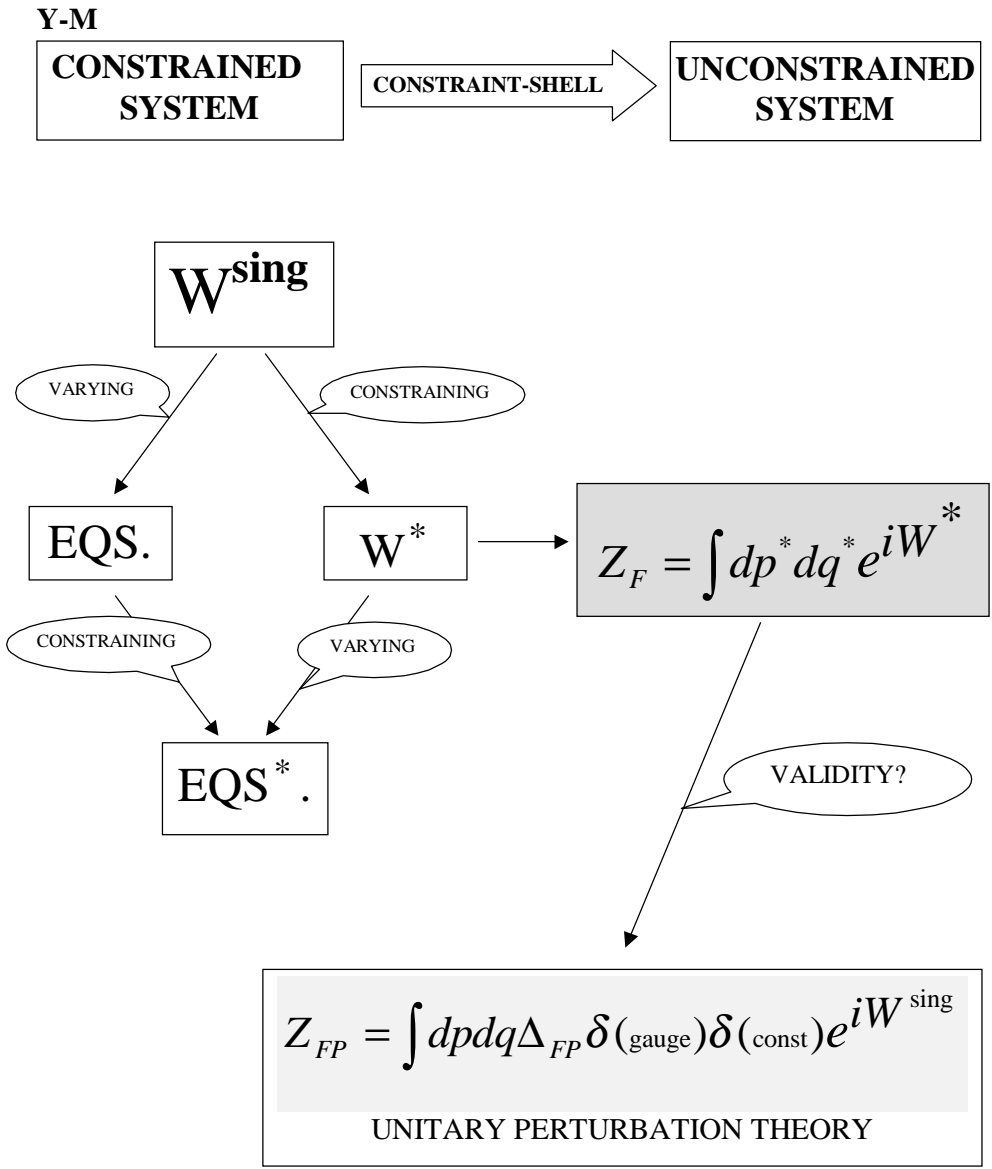
Fig. 2. An equivalent unconstrained system $W^*(p^*,q^*)$
can be obtained in the case
when the operations of varying and constraining commute with each other.
The next problem is to establish the range of validity of the standard
Faddeev-Popov (FP) integral.

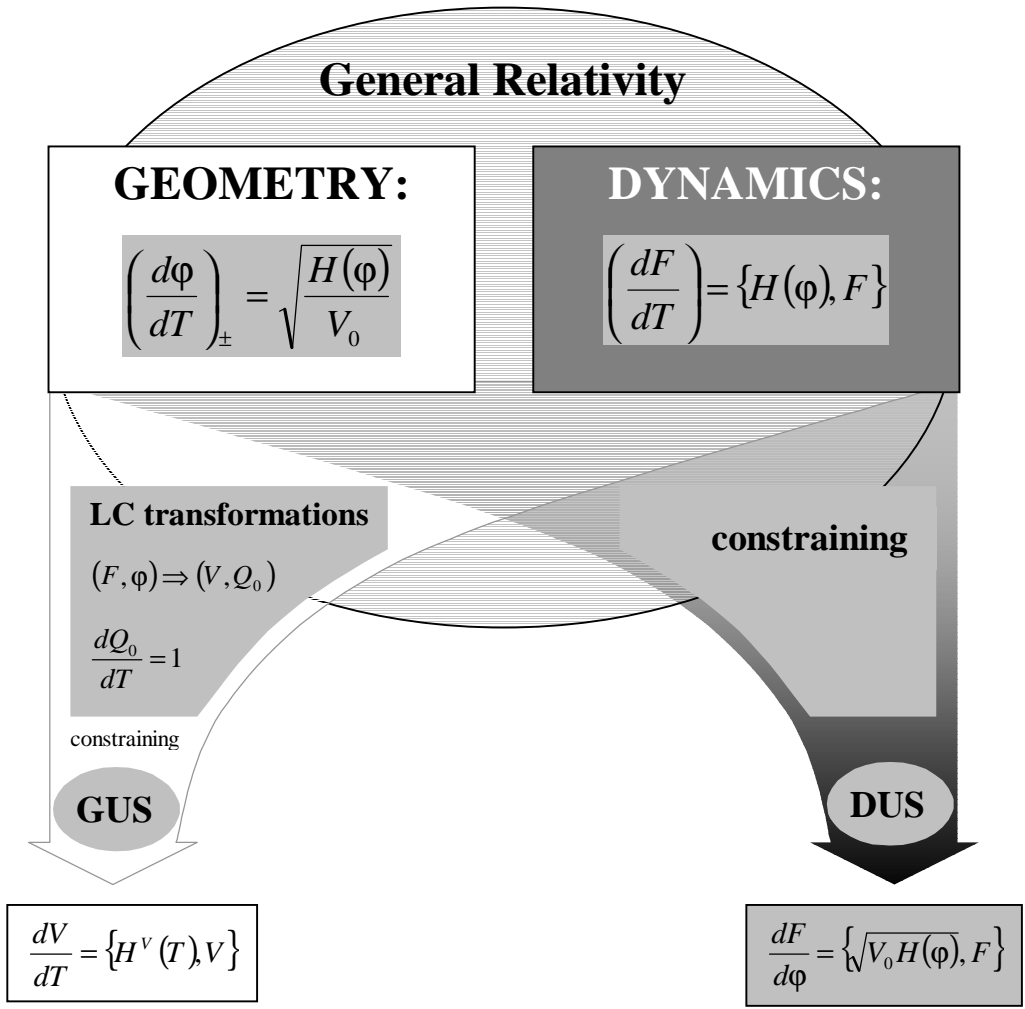
Fig. 3.
Reparametrization-invariant dynamics of General Relativity is
covered by the Dynamic Unconstrained Systems (DUS) and
the Geometric Unconstrained Systems (GUS)
connected by the Levi-Civita transformation of fields of MATTER
into the vacuum fields of initial data with respect to geometric TIME.

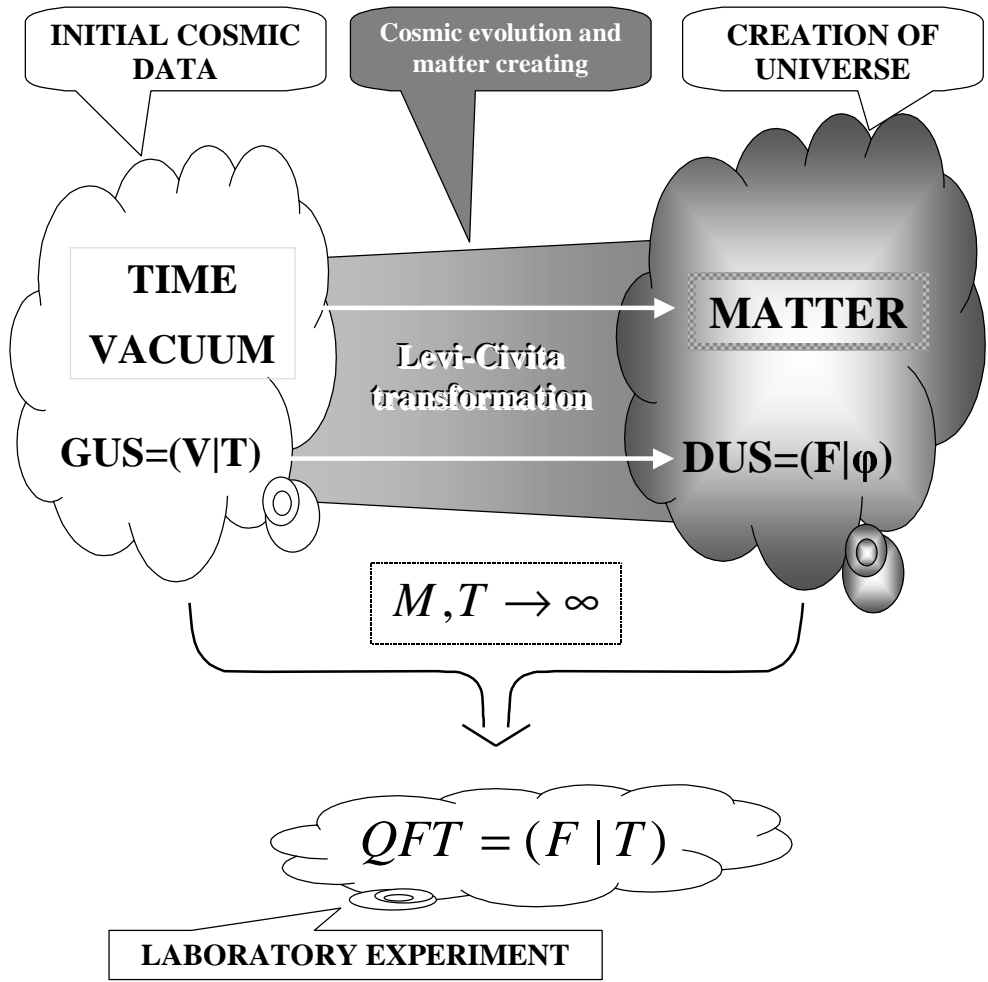
Fig. 4.
Reparametrization-invariant dynamics of General Relativity.
The Big Bang of Quantum Universe from point of view of Geometric
and Dynamic Unconstrained Systems connected by Levi-Civita
canonical transformations.


\begin{thebibliography}{}
\bibitem{fp2}
Faddeev L D and Popov V N 1973 {\it Us.Fiz.Nauk} {\bf 111} 427
\bibitem{dw}
DeWitt B S 1967  {\it Phys. Rev.} {\bf 160} 1113
\bibitem{gsw}
Green M B, Schwarz J H and Witten E 1986 {\it Superstring theory}
Cambridge University Press   Cambridge
( New York New Rochele Melbourne Sydney)
\bibitem{kpp}
Khvedelidze A, Palii Yu, Papoyan V and Pervushin V 1997
{\it Phys. Lett.} {\bf 402 B} 263
\bibitem{grg}
Gyngazov L N, Pawlowski M, Pervushin V N and
Smirichinski V I 1998, {\it Gen. Rel. and Grav.} {\bf 30} 1749
%
\bibitem{plb}
Pawlowski M,  Papoyan V V, Pervushin V N,
 Smirichinski V I 1998 {\it Phys. Lett. } {\bf 444B } 293
\bibitem{ps1}
Pervushin V N and Smirichinski V I 1999 {\it J.Phys.A: Math.Gen.}{\bf 32} 6191;
hep-th/9902013.
\bibitem{d1}
Dirac P A M 1958 {\it Proc.Roy.Soc.} {\bf A 246} 333
\bibitem{d2}
Dirac P A M 1964
{\it Lectures on Quantum Mechanics} (Belfer Graduate School of Science,
Yeshive University Press, New York)
\bibitem{kp}
Konopleva N P and Popov V N 1972 {\it Gauge fields} (Moscow, Atomizdat)
(in Russian)
\bibitem{gt}
Gitman D M and Tyutin I V  1986 {\it Canonical quantization
of fields  with constraints}
(Nauka, Moscow) (in Russian)
\bibitem{fp1} Faddeev L and  Popov V 1967
{\it Phys. Lett.} {\bf 25 B} 29
\bibitem{cj}
Dirac P A M 1955  {\it Can. J. Phys.} {\bf 33} 650
\bibitem{fad}
Faddeev L 1969
{\it Teor. Mat. Fiz.} {\bf 1} 3 (in Russian)
\bibitem{gip}
Gogilidze S, Ilieva N and Pervushin V 1999
{\it Int. J. Mod. Phys.} {\bf A 14} 3531;\\
Blashke D, Pervushin V and R\"opke G 1999 {\it Topological gauge-invariant
variables in QCD} UR-MPG-191/99 Preprint of Rostock Universit\"at,
hep-th/9909213
\bibitem{Y}
York J W (Jr.) 1971 {\it Phys. Rev. Lett.} {\bf 26} 1658\\
Kuchar K 1972 {\it J. Math. Phys.} {\bf 13} 768
\bibitem{kuchar}
Kuchar K 1992 {\it Time and Interpretations of Quantum Gravity }
in  (Proceedings of the Fourth  Canadian Conference
on General Relativity and Relativistic Astrophysics)
\bibitem{torre}
Torre C G 1991 {\it Class. Quantum Grav.}{\bf 8} 1895
\bibitem{yaf}
Pervushin V N and Smirichinski V I 1998 {\it Physics of Atomic Nuclei}
{\bf 61} 142
\bibitem{lc}
Levi-Civita T 1906 {\it Prace Mat.-Fiz.} {\bf 17} 1
\bibitem{sh}
Shanmugadhasan S 1973  {\it J. Math. Phys} {\bf  14\/} 677; \\
Lusanna L 1990 {\it Phys. Rep.} {\bf 185} 1
\bibitem{gkp}
Gogilidze S A, Khvedelidze A M and Pervushin V N 1996
{\it J. Math. Phys.} {\bf 37} 1760\\
Gogilidze S A, Khvedelidze A M and Pervushin V N 1996
{\it Phys. Rev. D} {\bf 53} 2160\\
Gogilidze S A, Khvedelidze A M and Pervushin V N 1999
{\it Phys.Particles and Nuclei} {\bf 30} 66
\bibitem{ADM}
Dirac P A M 1958 {\it Proc.Roy.Soc. A}{\bf 246} 333\\
Dirac P A M 1959 {\it Phys. Rev.} {\bf 114} 924\\
Arnovitt R, Deser S and Misner C W 1960 {\it Phys. Rev.} {\bf 117} 1595
\bibitem{vlad}
Vladimirov Yu S 1982 {\it Frame of references in
theory of gravitation} (Moscow, Energoizdat, in Russian)
\bibitem{L}
Lichnerowicz A 1944 {\it Journ. Math. Pures and Appl.} B{\bf 37} 23
\bibitem{bww}
Schweber S 1961 {\it An Introduction to Relativistic Quantum Field Theory}
(Row, Peterson and Co $\bullet$ Evanston, Ill., Elmsford, N.Y)
\bibitem{gmm}
Grib A A, Mamaev S G, Mostepanenko V M 1980
{\it Quantum effects in intensive external fields}
(Moscow, Atomizdat, in Russian)
\bibitem{b} Bogolubov N N 1947 {\it J.Phys.} {\bf 11} 23
\bibitem{M}
Misner C 1969 {\it Phys. Rev.} {\bf 186} 1319
\bibitem{pr}
Pawlowski M and Raczka R 1994 {\it Found. of Phys.} {\bf 24} 1305
\end{thebibliography}
\end{document}